# Quantum holographic encoding in a two-dimensional electron gas

Christopher R. Moon,[1] Laila S. Mattos,[1] Brian K. Foster,[2] Gabriel Zeltzer,[3] & Hari C. Manoharan[1*]

[1]Department of Physics, Stanford University, Stanford, California 94305, USA
[2]Department of Electrical Engineering, Stanford University, Stanford, CA 94305, USA
[3]Department of Applied Physics, Stanford University, Stanford, CA 94305, USA

*To whom correspondence should be addressed. E-mail: manoharan@stanford.edu

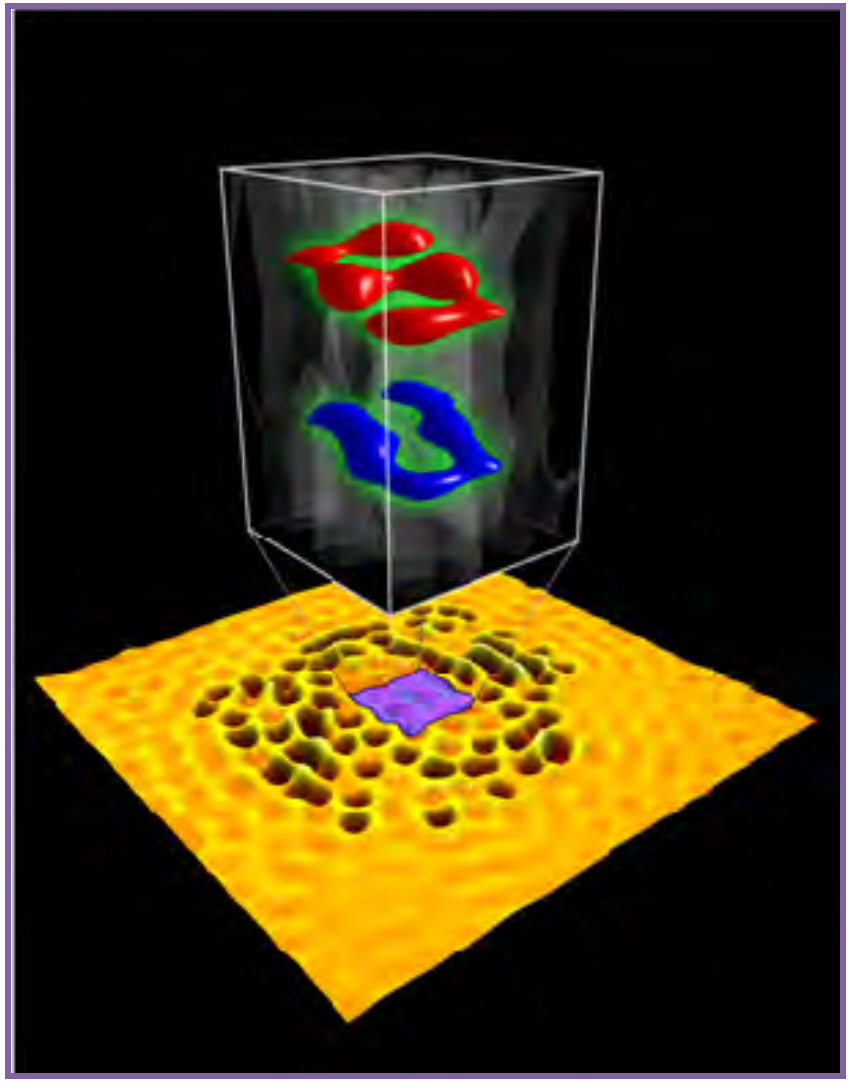

**The advent of bottom-up atomic manipulation[1] heralded a new horizon for attainable information density, as it allowed a bit of information to be represented by a single atom. The discrete spacing between atoms in condensed matter has thus set a rigid limit on the maximum possible information density. While modern technologies are still far from this scale, all theoretical downscaling of devices terminates at this spatial limit. Here, however, we break this barrier with electronic quantum encoding scaled to subatomic densities. We use atomic manipulation to first construct open nanostructures—'molecular holograms'— which in turn concentrate information into a medium free of lattice constraints: the quantum states of a two-dimensional degenerate Fermi gas of electrons. The information embedded in the holograms is transcoded at even smaller length scales into an atomically uniform area of a copper surface, where it is densely projected into both two spatial degrees of freedom and a third holographic dimension mapped to energy. In analogy to optical volume holography[2-4], this requires precise amplitude and phase engineering[5,6] of electron wavefunctions to assemble pages of information volumetrically. This data is read out by mapping the energy-resolved electron density of states with a scanning tunnelling microscope. As the projection and readout are both extremely near-field, and because we use native quantum states rather than an external beam[7], we are not limited by lensing or collimation and can create electronically projected objects with features as small as ~0.3 nm. These techniques reach unprecedented densities exceeding 20 bits/nm$^2$ and place tens of bits into a single fermionic state.**

In conventional holography, a three-dimensional (3D) object is stored in a two-dimensional (2D) hologram by interfering two coherent waves: an object beam, carrying information, and a simple reference beam. When the resulting interference pattern is later illuminated with a replica of the reference beam, the object is projected and the information can be retrieved (Fig. 1a). With fast digital techniques, the interference pattern can be simulated on a computer, with the hologram for a desired projection sent to a spatial light modulator. New developments in this arena include optical holographic data storage[2-4] (projecting data into a 3D storage medium), holographic optical tweezers[8,9] (creating precise 3D optical trapping potentials), and holovideo[10] (producing dynamic holographic projections in real time). Here we use a conjugate technique (Fig. 1b): we directly construct a scattering environment that is permanently illuminated by a continuous spectrum of electrons confined to a metal surface. The scattering geometry is arranged at the atomic scale such that a desired holographic projection results. However, this projection is not into three spatial dimensions, as in traditional optical or electron beam holography, but rather into 3D position-energy (**r**-$E$) space, where **r** = ($x, y$) is a planar position.

This new form of holography is achievable in any quantum 2D electron system where the local potential can be sufficiently controlled. In our implementation, we began with the nearly free gas of two-dimensional surface state electrons hosted by the (111) face of a Cu crystal[11]. Atomic manipulation enables confinement of these electrons into quantum corrals[12], permitting the study of lifetime effects[13], Kondo physics[14,15], single-atom gating[6], and quantum phase[5]. Curiously, however, limited attention has been paid to applications of electron waves in open geometries, other than the recognition that they can theoretically possess very small features[16]. Additionally motivated by existing sub-quantum detection schemes for both electronic charge[17] and magnetic flux[18], and new storage technology trends exploiting a third dimension[19], we sought to employ open, coherent quantum structures to encode data in a fashion scalable below the spatial limit of atomic condensed matter.

This quest resulted in molecular holograms (Fig. 1b) created by positioning CO molecules on the Cu(111) surface using a scanning tunnelling microscope (STM). Illuminated by the resident 2D electron gas (which is analogous to white light), these strictly planar holograms project electronic states into 3D position-energy space. Data can be encoded into this volumetric projection in various ways; here we project 2D pages of information into a readout area of the surface devoid of molecules. Page $i$ is a pattern of bits written at energy $E_i$ at positions **r** in the electronic local density of states (LDOS), $\rho(\mathbf{r}, E_i)$. Each bit is defined by whether $\rho(\mathbf{r}, E_i)$ is above or below a constant threshold value. A page is read with the STM as horizontal slices through the full 3D electronic object (Fig. 1b). In essence, information in the discrete molecule positions is concentrated into the central readout region, where we attempt to maximize information density. Holography is ideal for this task as it allows multiple pages to be stacked within the same physical space.

Coherent electron scattering in molecular holograms is not as simple as in closed quantum corrals, where electrons are well described by particle-in-a-box wavefunctions[6, 12]. Fortunately, the density of states can still be straightforwardly predicted with a multiple-scattering Green's function method[20, 21]. In sharp contrast, engineering a specific pattern in $\rho(\mathbf{r}, E)$—the encoding process—requires solving the difficult inverse problem. We used simulated annealing (Methods) to optimise the molecular hologram geometry. Related techniques have recently been applied[22] to find the theoretical shape of a quantum corral maximizing the quantum mirage[14]. Besides the molecular positions, free parameters are size and energy $E_i$ for an encoded page. Initially, we used electrons near the Fermi energy $E_F$ (at Fermi wavevector $k_F = \sqrt{2m^* E_F / \hbar^2}$ and wavelength $\lambda_F = 2\pi/k_F$) so they could be imaged at low sample bias $V$. The wavelength of $2k_F$ Friedel oscillations is $\lambda_F / 2 \approx 1.5$ nm on Cu(111), where $E_F = 0.45$ eV (corresponding to $V = 0$) and the effective mass $m^* = 0.38$ electron masses. For our baseline holography, we aimed to encode data with a linewidth of half of this wavelength, or $\lambda_F / 4 \approx 0.75$ nm. The quantum nature of the electrons provides a distinct advantage over holographic techniques based on free electrons or photons, because the Pauli exclusion principle guarantees a relatively large baseline energy, the Fermi energy, which is essentially built into the sample and accessible at zero bias.

Without loss of generality, we chose to encode pages whose readout (either high or low LDOS) would resemble letters of the alphabet. Starting with random molecular positions, we optimised the match between the theoretical $\rho(\mathbf{r}, E)$ and specific target bit patterns (Fig. 2, black-and-white images in right panels) for each page in the 2.9 × 4.3-nm$^2$ central readout area. Then, operating in ultrahigh vacuum at 4 K, we positioned the CO molecules according to the annealed designs, the first of which was for materializing the letter *S*. A constant-current (*I*) topograph of the assembled molecular hologram is shown in Fig. 2a. The centre of the topograph does not resemble the template image (Fig. 2b, upper right) because this topograph was acquired at $V = 10$ mV, whereas the target was −10 mV, demonstrating the energy sensitivity of the readout.

To decode the projection more precisely, we acquired constant-height $dI/dV$ maps (conductance maps—see Methods) of the image area. These are direct images of the electronic LDOS at specific energies. At $V = -17.5$ mV, a clear $S$ was revealed (Fig. 2b). If $\rho$ was measured solely at the intended bit locations of the target image, the encoded information was retrieved with perfect fidelity (Fig. 2b, lower right). This binary image persisted throughout a 20 mV range centred at $V = -17.5$ mV; that is, the $S$ is projected into every electronic state in a 2.9 nm × 4.3 nm × 20 mV region of **r**-$E$ space. The template image was 12 × 8 pixels so that the $S$ would image with fairly constant linewidth. However, this inter-pixel spacing (~0.4 nm) was too small for each pixel to be independently settable. Our experiments and simulations indicate that a 7 × 5 pixel array of this size allows full bit control, corresponding to a spacing of ~$\lambda_F / 4$ (0.75 nm). Thus, as a conservative estimate this $S$ page $\rho(\mathbf{r}, E_S) \equiv S(\mathbf{r})$ contained 35 bits of information; the areal information density was 2.8 bits/nm$^2$.

Using our nanoscale writing and encoding scheme, we also created a $U$ page at the same energy, with different molecular positions (Fig. 2c). The corresponding page $\rho(\mathbf{r}, E_U) \equiv U(\mathbf{r})$ is shown in Fig. 2d. We then constructed a hologram (Fig. 2e) projecting the $S$ and $U$ pages simultaneously. The result (Fig. 2f) is a single page $SU_{2D} = S(\mathbf{r}_1) + U(\mathbf{r}_2)$, containing both individual pages at the same energy but spatially offset by $|\mathbf{r}_1 - \mathbf{r}_2| = 3.6$ nm. (As in optical holography, even though the projection can be decomposed in this fashion, the $SU$ hologram is unrelated to the individual $S$ and $U$ holograms.) Each letter is ~3.0 × 4.7 nm$^2$ and combined they require twice the surface area as the $S$ or $U$ alone; the information density is essentially unchanged.

Moving beyond this baseline result, we encoded the $S$ and $U$ pages simultaneously in the same spatial area by embedding them volumetrically—a feat impossible with traditional surface writing. A topograph of the molecular hologram that accomplishes this is shown in Fig. 3a; since it was acquired at $V = -60$ mV, the centre of the image reflects the integration of the LDOS containing both pages. Conductance maps at energies $E_S = E_F - 18$ mV and $E_U = E_F - 45$ mV separate the density of states into the $S$ and $U$ (Fig. 3, b and c). We mapped the full electronic object by acquiring dense $dI/dV$ maps of the central readout region at many voltages. Fig. 3d displays a translucent surface of constant $dI/dV$ showing how the detected $\rho$ evolves through **r**-$E$ space. Slices through the four-dimensional dataset reveal that the information is contained in a truly volumetric function $SU_{3D} = S(\mathbf{r}, E_S) + U(\mathbf{r}, E_U)$, superposing the individual pages in the same space but offset in energy. Supplementary Video 1 shows the full readout of this data cube.

To quantify the positions of the encoded pages, we computed the normalized cross-correlation between each $dI/dV$ map and each template image. Fig. 3e shows surfaces of constant correlation surrounding the points where the correlation of each letter is greatest. This demonstrates that the $S$ and $U$ exist in the same region of space (localized to within 0.1 nm) but are separated by ~27 mV. Stacking the second page doubled the areal information density to 5.6 bits/nm$^2$, and highlights a novel scaling strategy for multiplexing information via volumetric holography.

The written features shown thus far have a characteristic size of ~0.8 nm, already much smaller than the wavelength of synchrotron-generated photons now used in state-of-the-art x-ray holography[23]. This can be even further reduced by using electrons at higher energies. An accurate measure of the wavevector content of electrons at a given energy is the spatial Fourier transform of $\rho$ (Fig. 4a, b). The result is a circular ring whose radial average (Fig. 4c) is peaked at wavevector $\kappa(E)$. As the voltage increases from $V = 10$ mV to 1.91 V, $\kappa$ grows by a factor of 2.8. In this way, electronic writing can be realized that surpasses spatial limits imposed by the coarseness of atoms.

Traditional writing scaled to its ultimate limit can be no smaller than atoms registered to a discrete underlying surface lattice[1]. Under this rubric, using eight Cu atoms manipulated with the STM tip, we created (Fig. 4f) the smallest atomic S possible on Cu(111). Atoms must be spaced by two lattice constants (0.51 nm); for example, at closer separations metal atoms bond to form inseparable clusters and cannot be placed arbitrarily. Electrons are subject to no such constraint. We created a significantly smaller $S(\mathbf{r})$ (Fig. 4e) using a molecular hologram (Fig. 4d) projecting into $V = 1.91$ V. As measured from the atom centres and the corresponding points in the density of states, the electronic S is ~1.24 nm in height, has linewidth ~0.3 nm, and occupies 50% of the area of its atomic counterpart. (Supplementary Figure 1 compares this to other nanoscale writing technologies and milestones.) The high-energy S page contains over 20 bits/nm$^2$, significantly exceeding the maximum atomic density of ~9 bits/nm$^2$. Note that this writing is even smaller than proposed[24] (presently technologically impossible) writing on a single highly excited Rydberg atom,[25] which would yield letters over 50 times taller than the $V = 1.91$-V projected S.

According to theory, a quantum state can encode any amount of information (at zero temperature), requiring only sufficiently high bandwidth and time in which to read it out[26]. In practice, only recently has progress been made towards encoding several bits into the shapes of bosonic single-photon wavefunctions[27], which has applications to quantum key distribution[28]. We have experimentally demonstrated that 35 bits can be permanently encoded into a time-independent fermionic state, and that two such states can be simultaneously prepared in the same area of space. In all variations shown, extending down to the subatomic regime, the encoded data is later retrieved at the bit level with 100% fidelity. Both the full theoretical and practical limits of this technique—the trade-off between information content (the number of pages and bits per page) and contrast (the number of measurements required per bit to overcome noise)—await further exploration.

*Methods*

**Forward scattering.** The multiple-scattering Green's function theory[20, 21] was used to predict the LDOS and hence our holographic projections. It requires the scattering phase shift for a molecule ($i\infty$ matches the data well at low biases) and the surface state dispersion relation (which is nonparabolic at high energies[29]). For $V \geq 1$ V, it was necessary to add a phase relaxation length $L$ to the free electron Green's function, such that $G_0(\mathbf{r}, \mathbf{r}', E) = G_{2D}(\mathbf{r}, \mathbf{r}', E) \exp(-|\mathbf{r} - \mathbf{r}'|/L)$, where $G_{2D}$ is the undamped propagator for a particle in two dimensions. For $V = 1.91$ V, we used $L = 5$ nm, which lessened but did not eliminate some mismatch to the data, and which is in agreement with the coherence length deduced from the momentum-space peak width (Fig. 4c).

**Inverse scattering.** CO molecules bond above the Cu atoms in the (111) surface, which form a triangular lattice with spacing $a = 0.255$ nm. A typical hologram requires optimising the positions of molecules restricted to a 15 × 15-nm$^2$ area, reserving the inner 4 × 5 nm$^2$ for the page readout area. For 80 molecules, there are ~3500 possible sites and $10^{167}$ ensembles. We used simulated annealing to optimize the molecular positions, which were initialised randomly. At each iteration, a random molecule was moved to an unoccupied site obeying design rules: sites having two occupied nearest neighbours were forbidden because they are physically unstable[30]. At higher voltages ($V \geq 1$ V) we found that all dimers became unstable so these were then forbidden as well.

The merit function for a projection (its match to the target bits) that we attempted to maximize was $f = \left[\min(W) - \max(B)\right] + b\left[\overline{W} - \overline{B}\right]$, where $W$ and $B$ are the sets of DOS computed in white and black pixels, respectively, and $\overline{W}$ denotes the mean of $W$. The first term represents the lowest contrast between high and low bits and the second term, the average contrast, encourages growth; we typically chose $b = 0.15-0.3$. In the simulated annealing for $SU_{3D}$ holography, we optimised the combined merit function $F = \min(f_{S,V_1}, f_{U,V_2})$, where the two $f$ terms represent the $S$ at $V_1$ and the $U$ at $V_2$. As a fictitious temperature $T$ was decreased, a change was accepted with probability $\exp(-\Delta f/T)$. We used a linear cooling schedule, starting at a $T$ that essentially randomised the molecular positions.

**Conductance mapping.** The $dI/dV$ maps were acquired open-loop: the tip was held at constant height (above the previously determined sample plane) and voltage as it was scanned. At each point, the response to a small AC voltage modulation (typically $V_{rms} = 4$ mV at 865 Hz) was recorded. By keeping the feedback loop disengaged, the $dI/dV$ signal was not scaled by changes in tip height due to the apparent topography; these maps faithfully reflect $\rho(\mathbf{r}, E)$.

**Acknowledgements.** This work was supported by US Office of Naval Research (YIP/PECASE N00014-02-1-0351), US National Science Foundation (CAREER DMR-0135122 & DMR-0804402), US Department of Energy (DE-AC02-76SF00515), and the Stanford-IBM Center for Probing the Nanoscale. We acknowledge the NDSEG program (C.R.M. & B.K.F.) and the Alfred P. Sloan Foundation (H.C.M.) for fellowship support. We thank L. Bozano, M. Brongersma, G. Burr, D. Eigler, G. Fiete, J. Kirtley, P. Kolchin, S. Harris, E. Heller, R. McGorty, V. Manoharan, J. Moon, J. Randel, S. Song, and Y. Yamamoto for discussions.

**Figure 1 | Electronic quantum holography concept. a,** In traditional optical holography, light shone on a 2D hologram projects a 3D object viewable by eye. **b**, In this work, two-dimensional quantum electrons illuminate coplanar holograms assembled with atomic manipulation. The analogous projection is an object of electron density of states, which is observed via scanning tunnelling microscopy. This projection is into one energetic and two spatial dimensions, spanning a 3D space into which information can be densely holographically encoded.

**Figure 2 | Holographic page encoding and readout.** First column: topographs of molecular holograms constructed with CO on the Cu(111) surface. Second column: high-resolution *dI/dV* maps ($V_{AC}$ = 4 mV rms; $I$ = 1 nA) of the resulting DOS at particular energies. Third column: encoded (top; binary image) and retrieved (bottom; *dI/dV* at pixel centres) information. **a, b,** *S* page (topo: 17 nm × 17 nm, *V* = 10 mV; *dI/dV:* 5 nm × 5 nm, -17.5 mV). **c, d,** *U* page (topo: 17 nm × 17 nm, -10 mV; *dI/dV*: 5 nm × 5 nm, -10 mV). **e, f,** Spatially offset *S* and *U* pages (topo: 25 nm × 25 nm, -10 mV; *dI/dV*: 8 nm × 8 nm, -10 mV). Central boxes on topographs mark the readout regions in which binary target data was encoded into the electronic DOS. The point-wise readout of the DOS retrieves the template images exactly.

**Figure 3 | Volumetric quantum holography. a,** Topograph (13.5 nm × 13.5 nm; $I$ = 1 nA; $V$ = -60 mV) of a molecular hologram that encodes two pages of data at different energies in the same region of space. **b,** A *dI/dV* map taken at *V* = -18 mV shows the *S* page. $V_{AC}$ = 4 mV rms. **c,** Measuring *dI/dV* at *V* = -45 mV retrieves the *U* page. The molecules are unchanged. Colour scales are the same as those in Fig. 2. **d,** By mapping *dI/dV* at many voltages between -80 mV and 10 mV, $\rho(\mathbf{r}, E)$ was measured throughout the readout region. A translucent surface of constant DOS is shown. Slices through this space at the appropriate energies reveal the *S* and *U* pages. **e,** The normalized cross-correlation of $\rho(\mathbf{r}, E)$ with each template image (insets) was computed as a function of **r** and *E*. Surfaces of constant correlation (at 98% of the global maximum) are shown for each page, confirming their locations in the information cube. Contours show the maximum correlation projected in each dimension (95%–99.9%).

**Figure 4 | Electronic versus atomic writing. a, b,** 2D Fourier transforms of electron standing waves (not shown) measured at *V* = 10 mV and 1.91 V, respectively. **c,** Their radial averages are each strongly peaked at a single wavevector $\kappa$ that increases with energy, allowing smaller features in the density of states. The weight surrounding the DC peak at $\kappa = 0$ is not plotted. **d,** Topograph of a molecular hologram (10 nm × 10 nm, $I$ = 1 nA, *V* = 10 mV) encoding an *S*. **e,** A *dI/dV* map (2.5 nm × 2.5 nm, $V_{AC}$ = 50 mV rms) taken at *V* = 1.91 V, a much higher energy than that of maps shown previously. The image was acquired inside the dashed readout region in **d**. The upper inset shows the encoded data and the bottom inset shows its retrieval. **f,** Topograph (same scale, $I$ = 1 nA, V = 10 mV) of the smallest *S* that can be written with atoms. Dots mark the centres of the eight atoms. The electronic *S* is significantly smaller, surpassing the ultimate scale of atoms on surfaces. The wavevector $\kappa$ corresponding to the minimum spacing between two atomic lines is shown as the green dotted line in **c**.

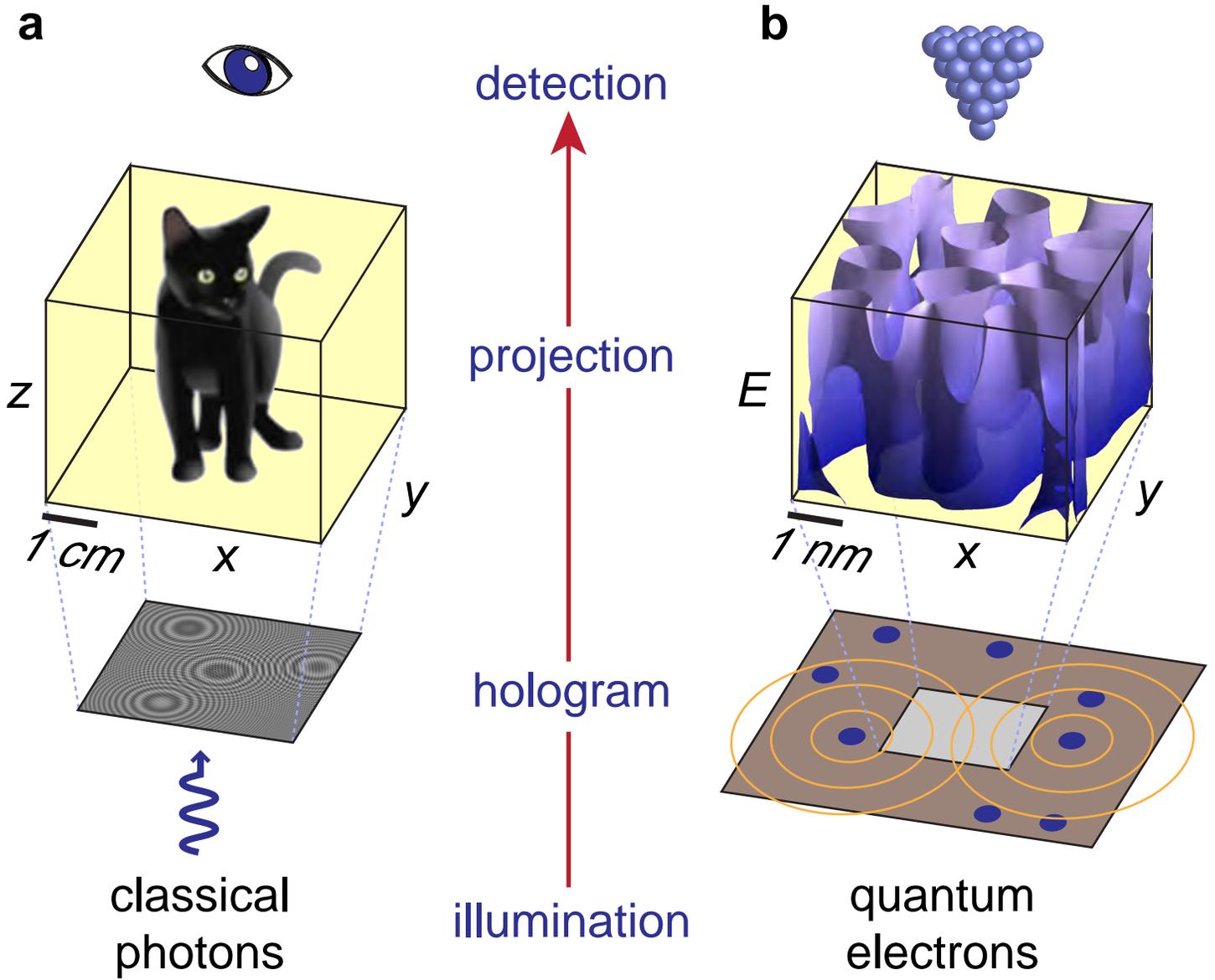

Figure 1

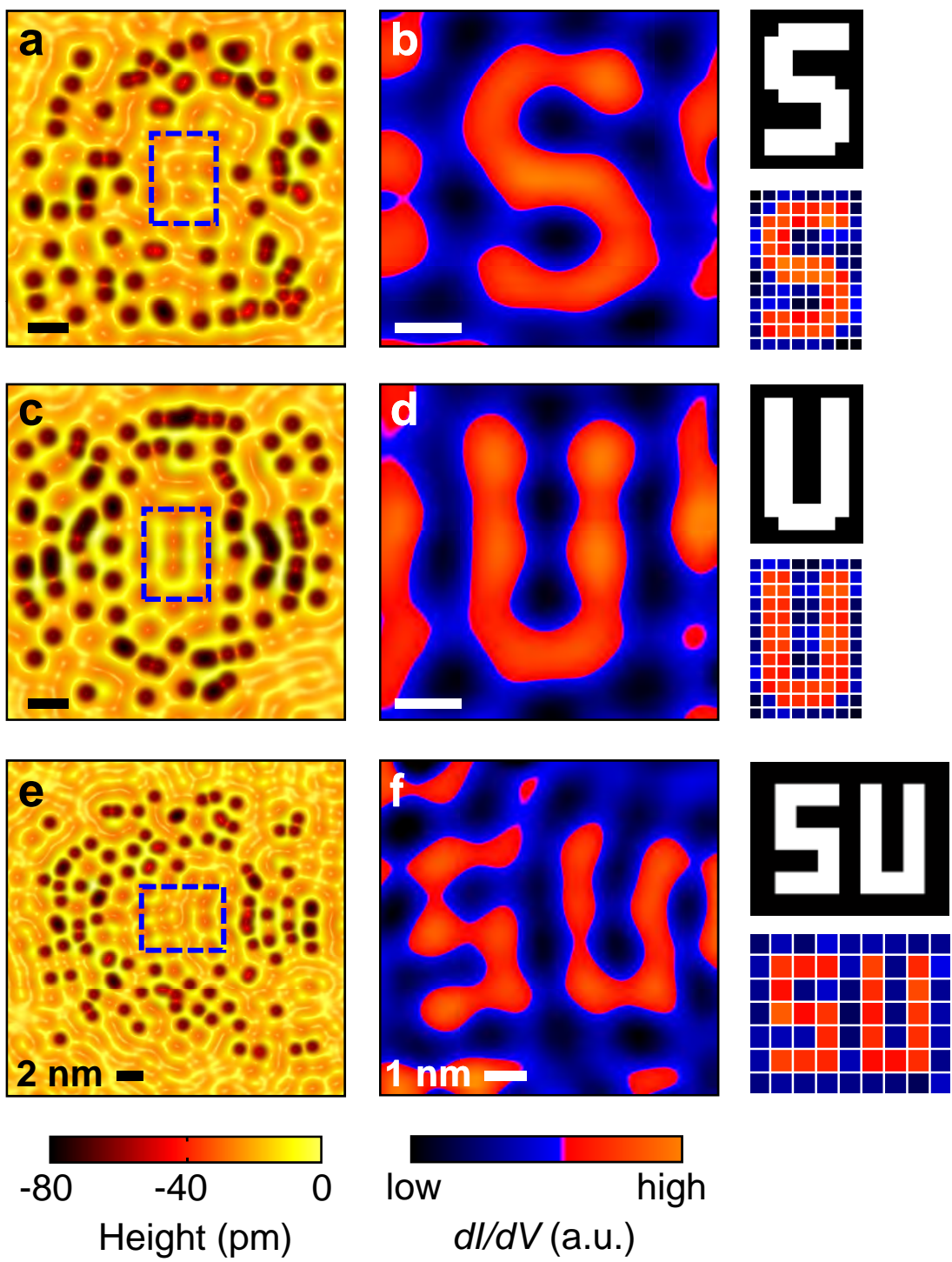

**Figure 2**

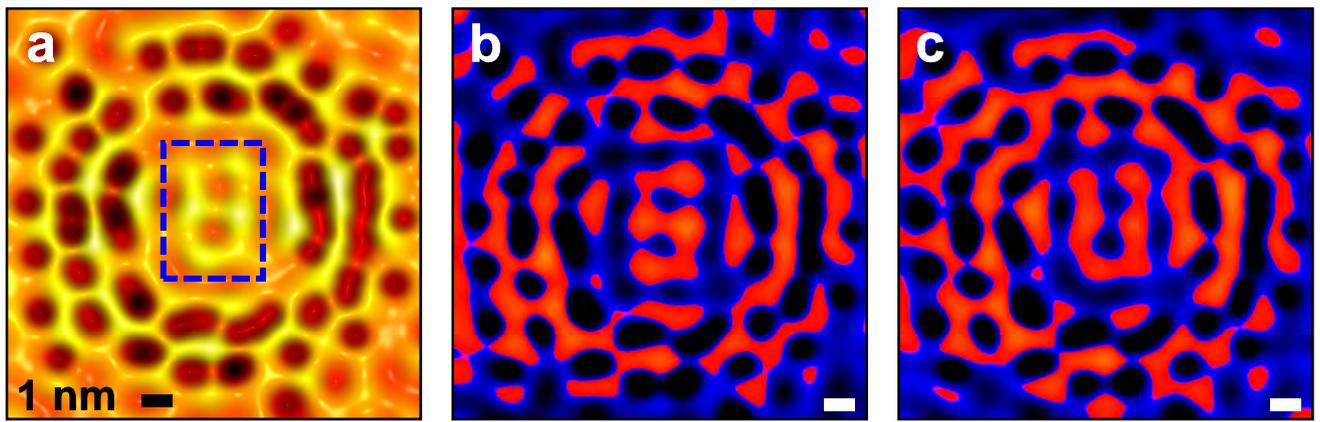
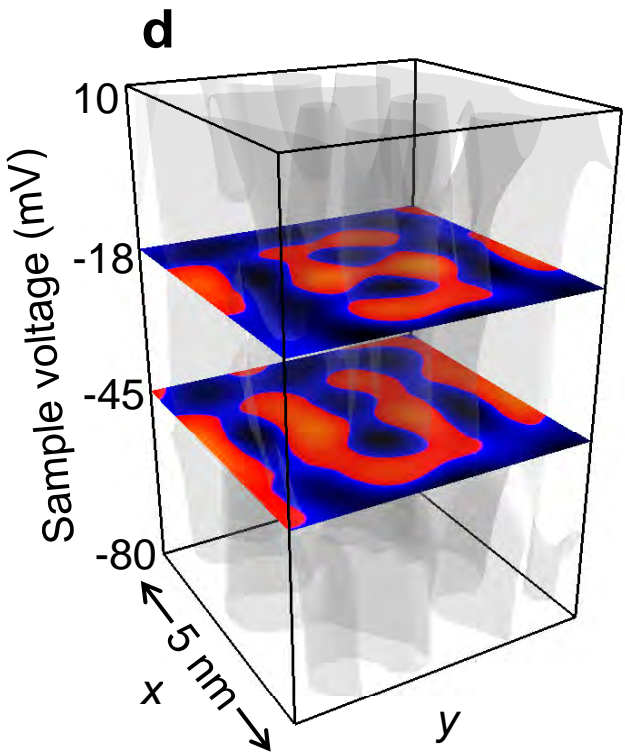
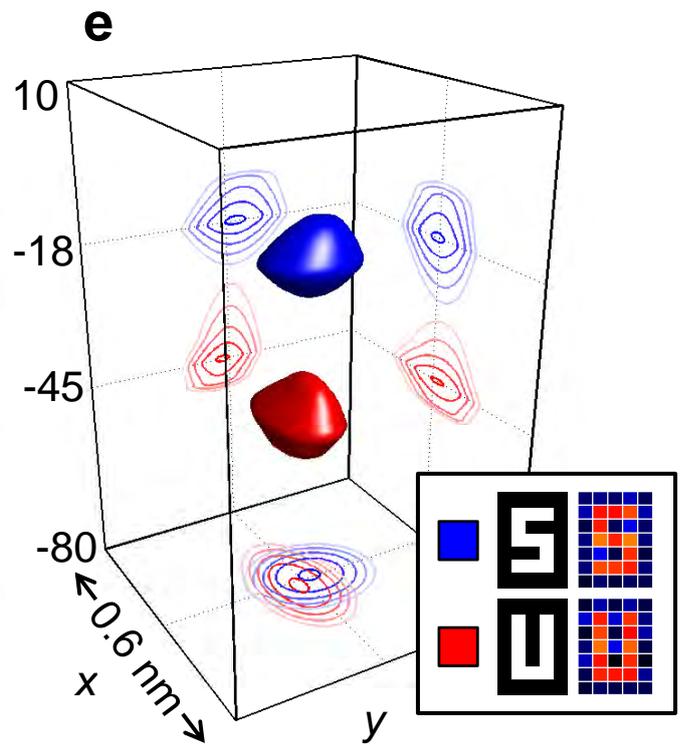

Electronic object projection

Target page correlation

**Figure 3**

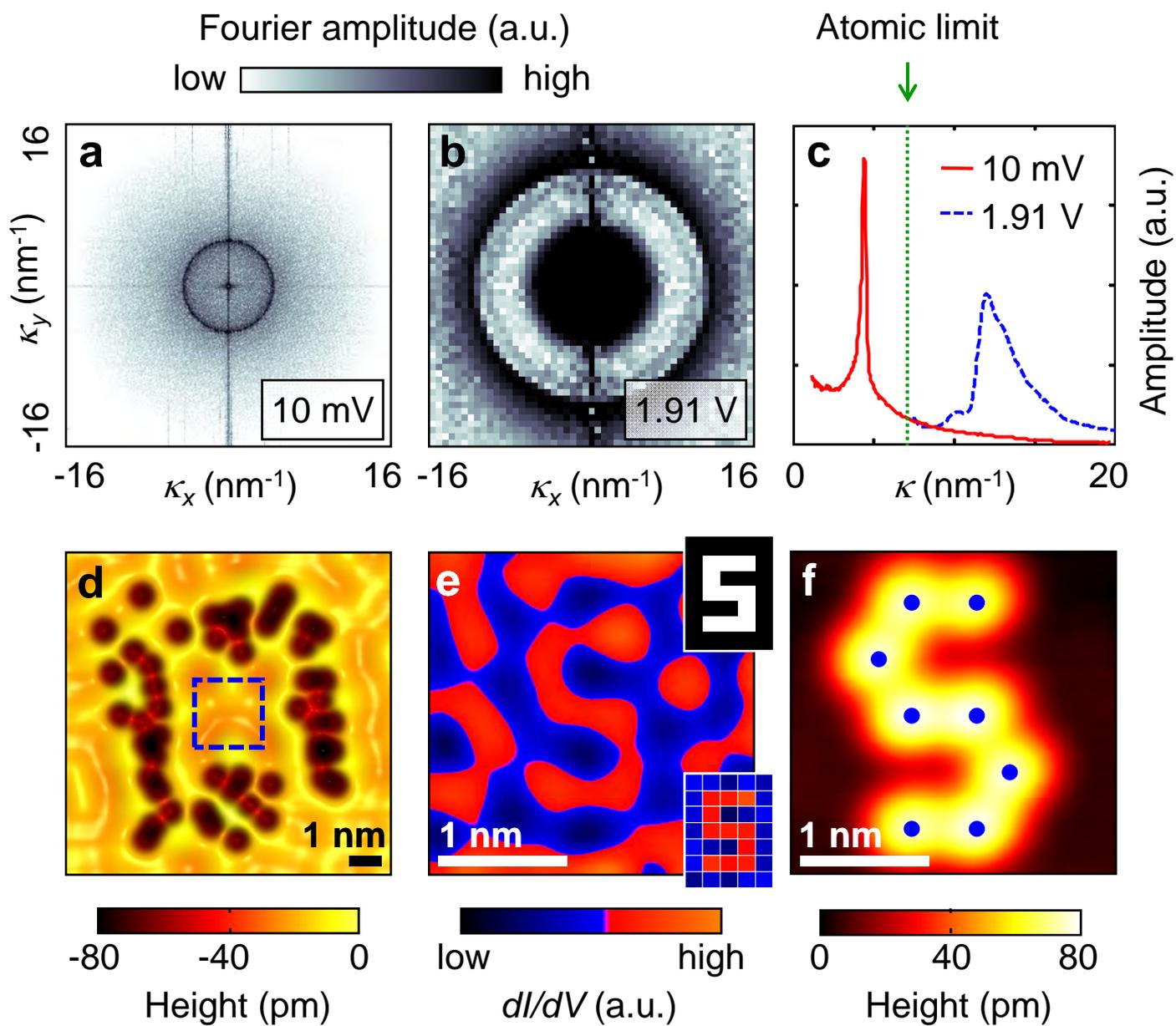

**Figure 4**

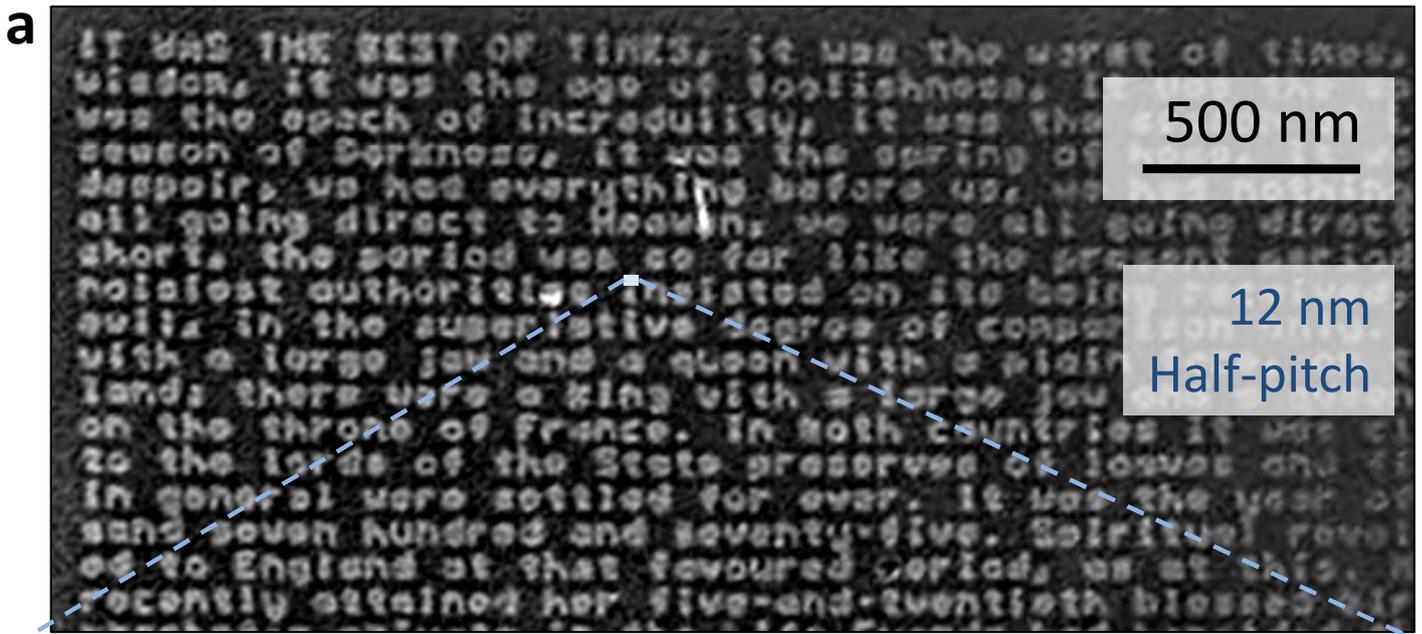

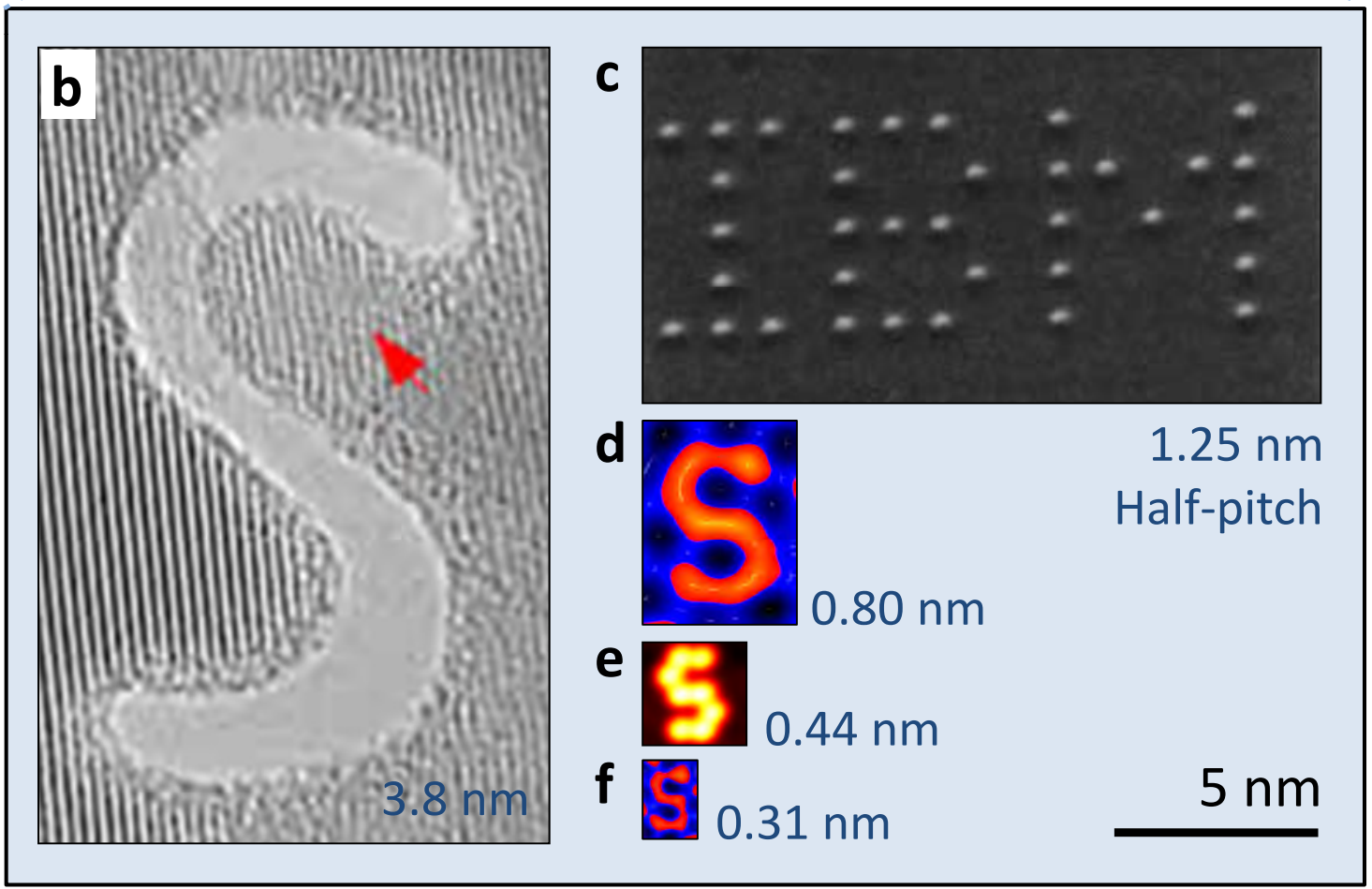

**Supplementary Figure 1 | Milestones in writing technologies. a,** Electron-beam lithography on $Si_3N_4$ that won the Feynman prize (adapted from ref. 31). **b,** Hole drilled through a Si nanowire using a high-energy electron beam (adapted from ref. 7). **c,** Individually positioned Xe atoms (adapted from ref. 1). *This work:* **d,** Baseline quantum electron holography (DOS at $E \approx E_F$). **e,** Maximally packed individually positioned metal atoms. **f,** Subatomic quantum electron holography (DOS at $E = E_F + 2$ eV). All items in the blue box share the same scale (black legend) and are arranged in order of decreasing feature half-pitch (blue legend). The bottom blue box is roughly the size of the dot in a letter *i* in the top text.

## Supplementary Notes

### Supplementary References

## Supplementary Video Legends

### Supplementary Video 1

This video shows the full readout of the holographic projection containing both the *S* and the *U* pages. Each slice of the four-dimensional dataset (density of states as a function of *x*, *y*, and energy) was acquired via *dI/dV* mapping. The *S* and *U* have been holographically embedded in this projection at sample voltages of -18 mV and -45 mV. A translucent surface of constant *dI/dV* is also shown. The color scale is the same as in Fig. 3.